\documentclass[11pt,letterpaper]{article}
\pdfoutput=1

\usepackage{jcappub}
\usepackage{amsmath}
\usepackage{amsfonts}
\usepackage{amssymb}
\usepackage{graphicx}
\usepackage{epsfig}

\def\be{\begin{equation}}
\def\ee{\end{equation}}
\def\bea{\begin{eqnarray}}
\def\eea{\end{eqnarray}}

\title{Characterizing unknown systematics in large scale structure surveys}

\author[a]{Nishant Agarwal,}
\author[a]{Shirley Ho,}
\author[b]{Adam D. Myers,}
\author[c]{Hee-Jong Seo,}
\author[d]{Ashley J. Ross,}
\author[e]{Neta Bahcall,}
\author[f]{Jonathan Brinkmann,}
\author[g]{Daniel J. Eisenstein,}
\author[h]{Demitri Muna,}
\author[i]{Nathalie Palanque-Delabrouille,}
\author[j]{Isabelle P\^{a}ris,}
\author[k]{Patrick Petitjean,}
\author[l,m]{Donald P. Schneider,}
\author[n,o]{Alina Streblyanska,}
\author[p]{Benjamin A. Weaver,}
\author[i]{and Christophe Y\`{e}che}

\affiliation[a]{McWilliams Center for Cosmology, Department of Physics, Carnegie Mellon University, \\ Pittsburgh, PA 15213, USA}
\affiliation[b]{Department of Physics and Astronomy, University of Wyoming, \\ Laramie, WY 82071, USA}
\affiliation[c]{Berkeley Center for Cosmological Physics, LBL and Department of Physics, University of California, \\ Berkeley, CA 94720, USA}
\affiliation[d]{Institute of Cosmology and Gravitation, University of Portsmouth, \\ Portsmouth, PO1 3FX, UK}
\affiliation[e]{Princeton University Observatory, \\ Peyton Hall, Princeton, NJ 08544, USA}
\affiliation[f]{Apache Point Observatory, \\ P.O. Box 59, Sunspot, NM 88349, USA}
\affiliation[g]{Harvard-Smithsonian Center for Astrophysics, \\ 60 Garden St., Cambridge, MA 02138, USA}
\affiliation[h]{Department of Astronomy, Ohio State University, \\ Columbus, OH 43210, USA}
\affiliation[i]{CEA, Centre de Saclay, Irfu/SPP, \\ F-91191 Gif-sur-Yvette, France}
\affiliation[j]{Departamento de Astronom\'{i}a, Universidad de Chile, \\ Casilla 36-D, Santiago, Chile}
\affiliation[k]{Universit\'{e} Paris 6 et CNRS, Institut d'Astrophysique de Paris, \\ 98bis blvd. Arago, 75014 Paris, France}
\affiliation[l]{Department of Astronomy and Astrophysics, Pennsylvania State University, \\ University Park, PA 16802, USA}
\affiliation[m]{Institute for Gravitation and the Cosmos, Pennsylvania State University, \\ University Park, PA 16802, USA}
\affiliation[n]{Instituto de Astrofisica de Canarias (IAC), \\ E-38200 La Laguna, Tenerife, Spain}
\affiliation[o]{Dept. Astrofisica, Universidad de La Laguna (ULL), \\ E-38206 La Laguna, Tenerife, Spain}
\affiliation[p]{Center for Cosmology and Particle Physics, New York University, \\ New York, NY 10003, USA}

\emailAdd{nishanta@andrew.cmu.edu}

\abstract{Photometric large scale structure (LSS) surveys probe the largest volumes in the Universe, but are inevitably limited by systematic uncertainties. Imperfect photometric calibration leads to biases in our measurements of the density fields of LSS tracers such as galaxies and quasars, and as a result in cosmological parameter estimation. Earlier studies have proposed using cross-correlations between different redshift slices or cross-correlations between different surveys to reduce the effects of such systematics. In this paper we develop a method to characterize unknown systematics. We demonstrate that while we do not have sufficient information to correct for unknown systematics in the data, we can obtain an estimate of their magnitude. We define a parameter to estimate contamination from unknown systematics using cross-correlations between different redshift slices and propose discarding bins in the angular power spectrum that lie outside a certain contamination tolerance level. We show that this method improves estimates of the bias using simulated data and further apply it to photometric luminous red galaxies in the Sloan Digital Sky Survey as a case study.}

\begin{document}

\maketitle


\section{Introduction}
\label{sec:intro}

Large scale structure (LSS) offers a complementary probe to the cosmic microwave background (CMB) to do precision cosmology. Owing to the large volume with full three-dimensional (3D) information that is available in LSS surveys, they offer the promise to provide competitive, or even better, constraints on cosmological parameters compared to CMB-based experiments, in the near future. For example, it is possible that the next level of improved constraints on primordial non-Gaussianity would come from LSS \cite{Fedeli:2009fj,Carbone:2010sb,Cunha:2010zz,Fedeli:2010ud,Giannantonio:2011ya,Pillepich:2011zz,Huang:2012mr,Carrasco:2012cv}.

Ideally, in order to maximize volume, one would like a full-sky spectroscopic LSS survey to very high redshift. The Baryon Oscillation Spectroscopic Survey (BOSS) \cite{Dawson:2012va} currently provides the largest volume of spectroscopic measurements \cite{Anderson:2012sa}, and next generation experiments such as the Dark Energy Spectroscopic Instrument (DESI) \cite{Schelgel:2011zz} and Euclid \cite{Cimatti:2009is} will increase the observed volume by an order of magnitude. However, redshift determination in photometric surveys is simpler and they invariably probe larger volumes. A large amount of Sloan Digital Sky Survey (SDSS) \cite{York:2000gk,Eisenstein:2011sa} data is based on photometry \cite{Ho:2012vy,Ho:2013lda}; with future surveys such as the Dark Energy Survey (DES) \cite{des} and the Large Synoptic Survey Telescope (LSST) \cite{Abell:2009aa} the amount of photometric data will virtually explode.

A major challenge in LSS photometric surveys, however, is understanding and accounting for various systematics in the data \cite{Ho:2012vy,Ross:2011cz,Huterer:2012zs,Pullen:2012rd,Giannantonio:2013uqa,Hernandez-Monteagudo:2013vwa,Leistedt:2013gfa}. Without properly correcting for sample systematics in the observed number density of objects, we cannot obtain an accurate measurement of the angular power spectrum of the LSS tracer, and in turn cannot extract cosmological information from it. Potential sources of contamination in the data include stellar obscuration, sky brightness, seeing variations, dust extinction, color offsets, and magnitude errors. While it has been possible to construct maps for these systematic fields and account for them in the data, what we will focus on in this paper is the effects of {\it unknown systematics}.\footnote{We define unknown systematics as any remaining systematics in the data after we have corrected for known systematics using given templates and assuming some model for the corrections. In addition to truly unaccounted sources of contamination in the data, these may include systematics due to inaccurate maps and/or an incomplete modeling of known systematic fields.}

We propose a method to use cross-correlations between different redshift slices to estimate contamination from unknown systematics in the auto-power spectra. Cosmologically, the cross-power between two LSS samples that have no overlap in redshift should be negligibly small.\footnote{Here we are not taking into account effects from lensing that could lead to a magnification bias for high redshift objects, as these effects will be small on the linear scales considered in this work \cite{Scranton:2005ci,Bernstein:2009bq,McQuinn:2013ib,dePutter:2013nha}. We are also assuming a negligible real-space contribution to the correlation signal on very large scales.} In practice, however, a real-observation detects a non-zero signal, either due to an overlap in redshift or due to common systematics \cite{Ho:2012vy,Pullen:2012rd}. We can, in fact, even estimate the cross-power theoretically using the overlap of the redshift distribution between different redshift slices. Subtracting the known systematic fields from the difference between the measured and theoretically expected cross-power yields the contribution from unknown systematics.

We introduce an ``unknown contamination coefficient'', which we denote ${\cal U}^{\alpha,\beta}_{\ell}$, between any two redshift slices $\alpha$ and $\beta$, $\ell$ being the multipole moment. This parameter can be thought of as approximately a product of the ratio of power due to unknown systematics to the observed power in each of the two redshift slices. We drop all $\ell$ bins in the auto-power spectra of both redshift slices $\alpha$ and $\beta$ for which ${\cal U}^{\alpha,\beta}_{\ell}$ is greater than some contamination threshold. We propose that only the remaining bins should be trusted to not be contaminated significantly with unknown systematics and consequently be used for cosmological parameter estimation.

The plan of this paper is as follows. We begin with a review of the theory and measurement of the angular power spectrum, for use later in the paper, in section \ref{sec:cl}. In section \ref{sec:systematics}, we describe our method to estimate the effects of unknown systematics and exclude significantly contaminated $\ell$ bins. We first test the method on simulated angular power spectra contaminated with unknown systematics and demonstrate that it leads to improved bias estimates, in section \ref{sec:sims}. In section \ref{sec:lrgs}, we then account for unknown systematics in the SDSS-III Data Release Eight (SDSS-III DR8) luminous red galaxies (LRGs) sample \cite{Aihara:2011sj,Ho:2012vy,Ross:2011cz}, as an example of application to real data. We conclude with a discussion in section \ref{sec:discussion}.

\section{Review of the angular power spectrum}
\label{sec:cl}

The power spectrum of the matter distribution in the Universe has two important features on scales relevant to current sky surveys. The first is a turn-over at $k \sim 0.01 h  \ {\rm Mpc}^{-1}$, which provides a measure of the size of the horizon at matter-radiation equality \cite{Dodelson:2003ft}. The second is oscillations caused by acoustic waves in the baryon-photon plasma in the early Universe, the scale of which is set by the sound horizon at hydrogen recombination at $z \sim 1000$ \cite{Peebles:1970ag,Sunyaev:1970er,Bond:1984fp,Holtzman:1989ki,Hu:1995en,Hu:1996vq,Eisenstein:1997ik}. Together these features offer the possibility of directly measuring the angular diameter distance as a function of redshift and thus constraining standard cosmological parameters \cite{Eisenstein:1998tu,Blake:2003rh,Hu:2003ti,Linder:2003ec,Matsubara:2002rf,Seo:2003pu,Matsubara:2004fr,White:2005tf,Dolney:2004va}.

Practically, it is simpler to carry out an imaging survey compared to a spectroscopic survey in a given amount of telescope time. We can therefore utilize large photometric surveys to more precisely measure the angular power spectrum, and compare these measurements to a theoretical 3D power spectrum by using the spectroscopic redshift distribution of the sample. In the following subsections we briefly review the theory and computation of the angular power spectrum.

\subsection{From the full sample distribution to the angular power spectrum}
\label{subsec:theorycl}

The overdensity of dark matter halos is related to the underlying matter overdensity through the bias factor $b_{1}$ (assuming a linear bias), $\delta_{\rm halo} = b_{1} \delta_{\rm matter}$. The  halo power spectrum for halos of mass $M$ at redshift $z$ is thus given by $P_{\rm halo}(M,k,z) = b_{1}^{2}(M,z) P_{\rm matter}(k,z)$.

The theoretical angular power spectrum can be calculated by projecting the full 3D power spectrum on the sky. Using the full Bessel integration on the largest scales and accounting for redshift space distortions as described, for example, in \cite{Padmanabhan:2006ku}, the angular power spectrum is given by
\bea
	C_{\ell} & = & C_{\ell}^{\rm gg} + C_{\ell}^{\rm gv} + C_{\ell}^{\rm vv} + a.
\label{eq:cl}
\eea
The superscripts ${\rm g}$ and ${\rm v}$ denote galaxy and velocity terms, respectively, and $a$ is an extra (constant) shot noise-like term that can be added to obtain a better fit to the non-linear power spectrum \cite{dePutter:2012sh}. The three contributions to the angular power spectrum above are given by the integrals \cite{Padmanabhan:2006ku}
\bea
	C_{\ell}^{\rm gg} & = & \frac{2}{\pi} \int {\rm d}\ln k \ k^{3} P_{\rm matter}(k,0) W_{\ell}^{2}(k), \\
	C_{\ell}^{\rm gv} & = & \frac{4}{\pi} \int {\rm d}\ln k \ k^{3} P_{\rm matter}(k,0) W_{\ell}(k) W_{\ell}^{r}(k), \\
	C_{\ell}^{\rm vv} & = & \frac{2}{\pi} \int {\rm d}\ln k \ k^{3} P_{\rm matter}(k,0) \left[ W_{\ell}^{r}(k) \right]^{2},
\eea
where the window functions can be calculated using
\bea
	W_{\ell}(k) & = & \int {\rm d}z \ b_{1} \frac{D(z)}{D(0)} \frac{{\rm d}N}{{\rm d}z} j_{\ell}(kr), \\
	W_{\ell}^{r}(k) & = & \int {\rm d}z \ \Omega_{m}^{0.56}(z) \frac{D(z)}{D(0)} \frac{{\rm d}N}{{\rm d}z} \Bigg[ \frac{2\ell^{2}+2\ell-1}{(2\ell-1)(2\ell+3)} j_{\ell}(kr) \nonumber \\
	& & \quad \quad - \ \frac{\ell(\ell-1)}{(2\ell-1)(2\ell+1)}j_{\ell-2}(kr) - \frac{(\ell+1)(\ell+2)}{(2\ell+1)(2\ell+3)}j_{\ell+2}(kr) \Bigg].
\eea
Here ${\rm d}N/{\rm d}z$ is the redshift distribution normalized to unity, $r(z)$ is the comoving distance, and $j_{\ell}(kr)$ is the $\ell^{\rm th}$ order spherical Bessel function. In the flat sky (large $\ell$) limit, one can switch to the Limber approximation \cite{Limber:1954zz}.

\subsection{Angular power spectrum estimation}
\label{subsec:oqe}

The angular power spectrum is usually calculated using the optimal quadratic estimator (OQE) method described in \cite{Seljak:1997ep,Tegmark:1997yq,Padmanabhan:2002yv,Padmanabhan:2006ku}. We start by parameterizing the power spectrum with $N$ step functions in $\ell$, $\tilde{C}_{\ell}^{i}$,
\bea
	C_{\ell} & = & \sum_{i} p_{i} \tilde{C}_{\ell}^{i},
\eea
where the $p_{i}$ are the parameters that determine the power spectrum. We form quadratic combinations of the data,
\bea
	q_{i} & = & \frac{1}{2} {\boldsymbol x}^{T} {\boldsymbol {\cal C}}_{i} {\boldsymbol {\cal C}}^{-1} {\boldsymbol {\cal C}}_{i} {\boldsymbol x},
\eea
where ${\boldsymbol x}$ is a vector of pixelized galaxy overdensities, ${\boldsymbol {\cal C}}$ is the covariance matrix of the data, and ${\boldsymbol {\cal C}}_{i}$ is the derivative of the covariance matrix with respect to $p_{i}$. The covariance matrix requires a prior power spectrum to account for cosmic variance; we estimate the prior by computing an estimate of the power spectrum with a flat prior and then iterating once. We also construct the Fisher matrix,
\bea
	F_{ij} & = & \frac{1}{2} {\rm tr} \left[ {\boldsymbol {\cal C}}_{i} {\boldsymbol {\cal C}}^{-1} {\boldsymbol {\cal C}}_{j} {\boldsymbol {\cal C}}^{-1} \right].
\eea
The power spectrum can then be estimated, $\hat{\boldsymbol p} = {\boldsymbol F}^{-1} {\boldsymbol q}$, with covariance matrix ${\boldsymbol F}^{-1}$.

\subsection{Markov-Chain Monte-Carlo procedure}
\label{subsec:mcmc}

Having discussed the theoretical and observed angular power spectra, we now describe our method to obtain the bias $b_{1}$ and the non-linear fitting parameter $a$ in each redshift slice.

We use a Markov-Chain Monte-Carlo (MCMC) approach to explore the available parameter space using a modified version of the widely-used package {\tt CosmoMC} \cite{Lewis:2002ah}. We calculate the linear matter power spectrum using the {\tt CAMB} code \cite{Lewis:1999bs} included in the {\tt CosmoMC} package, and apply the HaloFit prescription \cite{Smith:2002dz} to account for non-linear effects on the matter power spectrum. The resulting matter power spectrum is used in the equations of section \ref{subsec:theorycl} to calculate the theoretical angular power spectrum, which is then used in conjunction with the photometric power spectrum outlined in section \ref{subsec:oqe} to calculate the likelihood (assumed Gaussian), that is the input to the MCMC procedure,
\bea
	\chi^{2} & = & ({\boldsymbol d} - {\boldsymbol t})^{T} \ . \ {\boldsymbol {\cal C}}^{-1} \ . \ ({\boldsymbol d} - {\boldsymbol t}).
\eea
Here ${\boldsymbol d}$ is the data $C_{\ell}$ vector, ${\boldsymbol t}$ is the theory $C_{\ell}$ vector convolved with the full survey window function, and ${\boldsymbol {\cal C}}$ is, as before, the covariance matrix. Although the OQE used to calculate the data vector is designed to compute nearly anti-correlated power spectra across different multipole bins, it does retain a small contribution ($\lesssim 5\%$) from other multipole bins. We therefore convolve the theoretical spectrum with the full window function before calculating the likelihood. This step is especially important, for example, in models with non-zero primordial non-Gaussianity, since the power spectrum rises dramatically at low $\ell$ in these models \cite{Dalal:2007cu,Matarrese:2008nc,Slosar:2008hx}. Maximizing the likelihood in the full parameter space in our MCMC analysis provides constraints on the bias and the non-linear fitting parameter in each redshift slice.

For our analysis in sections \ref{sec:sims} and \ref{sec:lrgs} we use standard cosmological data, including the WMAP nine-year CMB data \cite{Hinshaw:2012fq,Bennett:2012fp} and  the ``Union 2'' supernova data set that includes 557 supernovae \cite{Amanullah:2010vv}, as our baseline model. The LSS data that we use is the observations of LRGs in the SDSS-III DR8 sample \cite{Aihara:2011sj,Ho:2012vy,Ross:2011cz}. As mentioned earlier, we only use photometric angular power spectra of LRGs. The underlying redshift distributions, however, are calculated using the BOSS spectroscopic redshifts of the same sample.

\section{Characterizing unknown systematics}
\label{sec:systematics}

Systematics limit the amount of information that can be extracted from any survey \cite{Huterer:2012zs,Pullen:2012rd}. The observed LSS tracer density field is likely to be contaminated with residual systematics, the most dominant sources being stellar contamination, sky brightness, and image quality/seeing variations. While it is possible (to some extent) to remove the effects of such {\it known} systematics, how do we account for {\it unknown} systematics? In this section we will describe a method to use cross-correlations between different redshift slices to selectively exclude certain $\ell$ bins that appear to have large unknown contamination. We propose that only the remaining bins should be used for cosmological parameter estimation.

\subsection{The method}

In the absence of strong evidence of non-linear effects of systematics on the observed density field, we adopt a linear relationship between systematics and the observed density \cite{Ho:2012vy,Ross:2011cz}. More general multiplicative errors have been considered in \cite{Gordon:2005ai,Huterer:2005ez,Huterer:2012zs,Hernandez-Monteagudo:2013vwa,Anderson:2013zyy}. We find that systematics in the SDSS galaxy data that we consider are well-modeled with additive errors and adopt this simple parameterization here. We therefore write the following expression,
\bea
	\delta_{g,{\rm obs}}^{\alpha}(\ell,m) & = & \delta_{g,{\rm true}}^{\alpha}(\ell,m) + \sum_{i=1}^{N_{\rm sys}} \epsilon_{i}^{\alpha}(\ell) \delta_{i}(\ell,m) + u^{\alpha}(\ell,m),
\eea
where $\delta_{g,{\rm obs}}^{\alpha}(\ell,m)$ and $\delta_{g,{\rm true}}^{\alpha}(\ell,m)$ are the observed and true tracer density fields in the $\alpha^{\rm th}$ redshift slice, $\delta_{i}(\ell,m)$ is the contribution of the $i^{\rm th}$ systematic (with a total of $N_{\rm sys}$ systematics) to the density map, and $\epsilon_{i}^{\alpha}(\ell)$ (assumed independent of $m$) is a weight factor that characterizes the effect of the $i^{\rm th}$ systematic. We further parameterize the effect of any unknown systematics as $u^{\alpha}(\ell,m)$. The above equation holds in each $\ell$ bin, with the observed and true angular auto-power spectrum in each redshift slice defined as $C_{\ell, {\rm obs}}^{\alpha,\alpha} \equiv \big\langle \delta_{g,{\rm obs}}^{\alpha}(\ell,m) \delta_{g,{\rm obs}}^{\alpha}(\ell,m) \big\rangle$ and $C_{\ell, {\rm true}}^{\alpha,\alpha} \equiv \left\langle \delta_{g,{\rm true}}^{\alpha}(\ell,m) \delta_{g,{\rm true}}^{\alpha}(\ell,m) \right\rangle$. Here we have assumed isotropy as there is no evidence for anisotropy in the LSS power spectrum \cite{Pullen:2010zy}.

Let us assume that the true density field is not correlated with any of the systematics, so that $\left\langle \delta_{g,{\rm true}}^{\alpha}(\ell,m) \delta_{i}(\ell,m) \right\rangle = 0$ and $\left\langle \delta_{g,{\rm true}}^{\alpha}(\ell,m) u^{\beta}(\ell,m) \right\rangle = 0$. We can measure the observed auto-power spectra $C_{\ell, {\rm obs}}^{\alpha,\alpha}$, the cross-correlations between the observed density map and systematics $\big\langle \delta_{g,{\rm obs}}^{\alpha}(\ell,m) \delta_{i}(\ell,m) \big\rangle$, and the correlations amongst various systematics $\langle \delta_{i}(\ell,m) \delta_{j}(\ell,m) \rangle$. We can now write the following set of $N_{\rm sys}$ equations in each $\ell$ bin of the $\alpha^{\rm th}$ redshift slice,
\bea
	& & \left\langle \delta_{g,{\rm obs}}^{\alpha}(\ell,m) \delta_{j}(\ell,m) \right\rangle \nonumber \\
	& & \quad \quad \quad = \ \sum_{i = 1}^{N_{\rm sys}} \epsilon_{i}^{\alpha}(\ell) \langle \delta_{i}(\ell,m) \delta_{j}(\ell,m) \rangle + \langle u^{\alpha}(\ell,m) \delta_{j}(\ell,m) \rangle, \quad j = 1, \ldots, N_{\rm sys}.
\label{eq:solveep}
\eea
The unknown quantities in the above equations are the weights $\epsilon_{i}^{\alpha}(\ell)$ and the correlations $\langle u^{\alpha}(\ell,m)$ $\delta_{j}(\ell,m) \rangle$. We do not have a sufficient number of equations to solve for all of these quantities (we have $N_{\rm sys}$ equations and $2N_{\rm sys}$ unknowns). We investigated the use of cross-correlations between different redshift slices to solve for the remaining unknowns, but this approach failed, as we will now demonstrate.

We can write the following equation connecting the true and observed angular auto- or cross-power spectra,
\bea
	C_{\ell, {\rm true}}^{\alpha,\beta} & = & C_{\ell, {\rm obs}}^{\alpha,\beta} - \sum_{i,j = 1}^{N_{\rm sys}} \epsilon_{i}^{\alpha}(\ell) \epsilon_{j}^{\beta}(\ell) \langle \delta_{i}(\ell,m) \delta_{j}(\ell,m) \rangle - U^{\alpha,\beta}_{\ell},
\label{eq:crosspower}
\eea
where $U^{\alpha,\beta}_{\ell} = \sum_{i = 1}^{N_{\rm sys}} \Big( \epsilon_{i}^{\alpha}(\ell) \langle \delta_{i}(\ell,m) u^{\beta}(\ell,m) \rangle + \epsilon_{i}^{\beta}(\ell) \langle \delta_{i}(\ell,m) u^{\alpha}(\ell,m) \rangle \Big) + \langle u^{\alpha}(\ell,m) u^{\beta}(\ell,m) \rangle$. Although we do not know a priori the true auto-power spectra $C_{\ell, {\rm true}}^{\alpha,\alpha}$, we can estimate the true cross-power spectra $C_{\ell, {\rm true}}^{\alpha,\beta}$ ($\alpha \neq \beta$) theoretically using the equations given in section \ref{subsec:theorycl}. Here we assume that the true cross-power closely follows background cosmology, although, strictly speaking, this is correct only if the overlap in the redshift distribution between the two slices is small. We need the bias in each of the two redshift slices $\alpha$ and $\beta$ --- let us assume that we know this --- and the cross-redshift distribution, which we can calculate as the overlap of that in the two redshift slices. Then the only remaining unknowns in these equations are the weights $\epsilon_{i}^{\alpha}(\ell)$, the correlations $\langle u^{\alpha}(\ell,m) \delta_{j}(\ell,m) \rangle$, and additionally the correlations $\langle u^{\alpha}(\ell,m) u^{\beta}(\ell,m) \rangle$. One can easily check that we still lack a sufficient number of equations to solve for all unknown quantities. Therefore we can {\it not} correct for unknown systematics in the data. We {\it can}, however, estimate their contribution and decide to drop those $\ell$ bins that are dominated by unknown systematics.

In order to achieve this goal, we first obtain a zeroth order estimate of the weights $\epsilon_{i}^{\alpha}(\ell)$ by solving eq.\ (\ref{eq:solveep}) in each $\ell$ bin under the assumption that $u^{\alpha}(\ell,m) = 0$. We then obtain the systematics-corrected auto-power spectrum in each redshift slice (again under the assumption that $u^{\alpha}(\ell,m) = 0$) using 
\bea
	C_{\ell, {\rm true}}^{\alpha,\alpha} & = & C_{\ell, {\rm obs}}^{\alpha,\alpha} - \sum_{i,j = 1}^{N_{\rm sys}} \epsilon_{i}^{\alpha}(\ell) \epsilon_{j}^{\alpha}(\ell) \langle \delta_{i}(\ell,m) \delta_{j}(\ell,m) \rangle.
\label{eq:autopower}
\eea
We now perform an MCMC procedure using the above auto-power spectra to fit for the background cosmology, the bias, and the non-linear fitting parameter $a$ (introduced in eq.\ (\ref{eq:cl})) in each redshift slice. With these parameters in hand, we obtain a theoretical estimate of $C_{\ell, {\rm true}}^{\alpha,\beta}$ ($\alpha \neq \beta$) as described earlier. We also include the $a$ parameter in the cross-power --- ideally this should be zero, so we add the maximum value of $\sqrt{|a^{\alpha}| \ |a^{\beta}|}$ (where $a^{\alpha}$ and $a^{\beta}$ correspond to values of $a$ for redshift slices $\alpha$ and $\beta$) only in the cross-power of neighboring redshift slices. Finally, using eq.\ (\ref{eq:crosspower}) we solve for $U^{\alpha,\beta}_{\ell}$ ($\alpha \neq \beta$).\footnote{When calculating $U^{\alpha,\beta}_{\ell}$, we do not convolve the theoretical cross-power with the survey window function.}

We now define an unknown contamination coefficient as
\bea
	{\cal U}^{\alpha,\beta}_{\ell} & = & \frac{\left( U^{\alpha,\beta}_{\ell} \right)^{2}}{C_{\ell, {\rm obs}}^{\alpha,\alpha} C_{\ell, {\rm obs}}^{\beta,\beta}}, \quad \alpha \neq \beta,
\label{eq:defineul}
\eea
and compare it to the quantity
\bea
	n^{2} \frac{\sigma \big( C_{\ell, {\rm obs}}^{\alpha,\alpha} \big) \sigma \big( C_{\ell, {\rm obs}}^{\beta,\beta} \big)}{C_{\ell, {\rm obs}}^{\alpha,\alpha} C_{\ell, {\rm obs}}^{\beta,\beta}},
\label{eq:usyssigma}
\eea
where $n$ refers to the sigma tolerance level. We drop all $\ell$ bins in the auto-power of each redshift slice $\alpha$ and $\beta$ for which ${\cal U}^{\alpha,\beta}_{\ell}$ is greater than the quantity in eq.\ (\ref{eq:usyssigma}).\footnote{Clipping the cross-power at $n\sigma$ will in general introduce a bias in the variance of each auto-power spectrum. We checked that this effect is small for two Gaussian random distributions, introducing a percent level bias for a $2\sigma$-cut. We ignore this effect here.} For example, we can choose $n = 1$ and perform a $1\sigma$-cut on unknown systematics. The motivation for this definition is that it provides some idea of unknown contamination in units of the error in the observed auto-power. This is not a precise cut, however, as there is an error associated with the cross-power itself as well. Nevertheless, it does serve as a uniform measure to quantify the level of unknown contamination. After performing this cut, we use the remaining bins in the auto-power spectra to fit for the background cosmology, the bias and the $a$ parameter in each redshift slice. If the new bias values lie within one sigma of the previous estimates of the bias then these are the final values to use. If not, then we repeat this analysis until the bias lies within one sigma of the previous iteration.
\\ \\
{\bf Summary.} To summarize, we use the following algorithm to characterize unknown systematics ---
\begin{enumerate}
	\item Correct the observed auto-power spectra for known systematics, assuming no unknown contamination, using eq.\ (\ref{eq:solveep}) with $u^{\alpha}(\ell) = 0$ and eq.\ (\ref{eq:autopower}).
	\item Perform an MCMC analysis over background cosmological parameters, the bias, and the non-linear fitting parameter $a$ in each redshift slice.
	\item Use best-fit values of the bias and $a$ parameters to obtain theoretical cross-power spectra. Compare the resulting true cross-power with the observed cross-power using eq.\ (\ref{eq:crosspower}) to obtain $U_{\ell}^{\alpha,\beta}$, and subsequently obtain ${\cal U}_{\ell}^{\alpha,\beta}$ defined in eq.\ (\ref{eq:defineul}).
	\item Perform an $n\sigma$-cut on unknown systematics by excluding all $\ell$ bins in each auto-power spectrum for which ${\cal U}_{\ell}^{\alpha,\beta}$ with any other redshift slice is greater than the quantity in eq.\ (\ref{eq:usyssigma}).
	\item Using remaining bins in the auto-power spectra, perform an MCMC analysis over background cosmological parameters, the bias, and the parameter $a$ in each redshift slice, to obtain new estimates of various parameters.
\end{enumerate}
We repeat steps 3, 4, and 5 until the new bias values in each redshift slice lie within one sigma of the previous iteration.

\subsection{Estimating the covariance matrix}

In the above discussion we have not yet defined the errors on the angular auto-power spectra and the full covariance matrix between different redshift slices that will be needed for the MCMC fitting. In the absence of a complete understanding of mock systematic fields, it is unlikely that we can obtain optimal errors on systematic corrections within the scope of this analysis. We therefore adopt a simplistic model to estimate the covariance of the systematics-corrected power spectra, similar to \cite{Ho:2012vy,Ross:2011cz}, which we describe below.

Assuming Gaussianity of the fields involved, the covariance matrix has a simple structure of non-zero diagonal elements, with entries for multipoles $\ell^{\alpha} = \ell^{\beta}$ for redshift slices $\alpha$ and $\beta$. These covariances are given by
\bea
	\sigma^{2} \big( C_{\ell}^{\alpha,\alpha} \big) & = & \frac{2}{f_{\rm sky}(2\ell + 1)} \left( C_{\ell,{\rm smooth}}^{\alpha,\alpha} + N_{{\rm shot},\alpha} \right)^{2}, \\
	\sigma^{2} \big( C_{\ell}^{\alpha,\alpha} C_{\ell}^{\beta,\beta} \big) & = & \frac{2}{f_{\rm sky}(2\ell + 1)} \left( C_{\ell,{\rm smooth}}^{\alpha,\beta} \right)^{2},
\eea
where $\sigma^{2} \big( C_{\ell}^{\alpha,\alpha} \big)$ is the (diagonal) variance in the auto-power of redshift slice $\alpha$ and $\sigma^{2}\big( C_{\ell}^{\alpha,\alpha}$ $C_{\ell}^{\beta,\beta} \big)$ is the (diagonal) cross-covariance between redshift slices $\alpha$ and $\beta$. Note that the cross-covariance between two redshift slices is different from the variance in the cross-power between the two redshift slices, which is given by eqs.\ (\ref{eq:covcross1}) and (\ref{eq:covcross2}) in the next section. Here $f_{\rm sky}$ is the fraction of the sky observed and the shot noise is given by $N_{\rm shot} = f_{\rm sky} \times 4\pi/N_{\rm sample}$, where $N_{\rm sample}$ is the effective number of objects observed (e.g.\ galaxies, after weighting by the probability that an observed object is a galaxy). The shot noise term only contributes to the variance of the auto-power spectra. The subscript `smooth' denotes a smooth fit to the observed auto- or cross-power spectra. Also, the above equations assume a unit width for each $\ell$ bin, i.e. $\Delta \ell = 1$.

We modify the above set of equations by adding the correctional power due to systematics \cite{Ho:2012vy,Ross:2011cz},
\bea
	\sigma^{2} \big( C_{\ell}^{\alpha,\alpha} \big) & = & a^{2}_{\rm fac} \frac{2}{f_{\rm sky} \sum_{\ell = \ell_{\rm min}}^{\ell_{\rm max} - 1}(2\ell + 1)} \left( \sqrt{ \left( C_{\ell,{\rm smooth}}^{\alpha,\alpha} \right)^{2} + \Big( \Delta C_{\ell}^{\alpha,\alpha} \Big)^{2} } + N_{{\rm shot},\alpha} \right)^{2}, \quad \quad
\label{eq:cov1} \\
	\sigma^{2} \big( C_{\ell}^{\alpha,\alpha} C_{\ell}^{\beta,\beta} \big) & = & a^{2}_{\rm fac} \frac{2}{f_{\rm sky} \sum_{\ell = \ell_{\rm min}}^{\ell_{\rm max} - 1}(2\ell + 1)} \bigg( \left( C_{\ell,{\rm smooth}}^{\alpha,\beta} \right)^{2} + \left( \Delta C_{\ell}^{\alpha,\beta} \right)^{2} \bigg),
\label{eq:cov2}
\eea
where $\Delta C_{\ell}^{\alpha,\alpha} = \sum_{i,j = 1}^{N_{\rm sys}} \epsilon_{i}^{\alpha}(\ell) \epsilon_{j}^{\alpha}(\ell) \langle \delta_{i}(\ell,m) \delta_{j}(\ell,m) \rangle$ and $\Delta C_{\ell}^{\alpha,\beta} = \sum_{i,j = 1}^{N_{\rm sys}} \epsilon_{i}^{\alpha}(\ell) \epsilon_{j}^{\beta}(\ell) \langle \delta_{i}(\ell,m)$ $\delta_{j}(\ell,m) \rangle$. We have also taken into account the fact that $\Delta \ell \neq 1$, with the total number of modes in each $\ell$ bin coming from the sum of modes in $\ell_{\rm min} \leq \ell < \ell_{\rm max}$. Further, we have boosted the diagonal error with an empirical factor of $a_{\rm fac}$ to account for the fact that the Gaussian approximation is not perfect and neighboring $\ell$ bins contribute to the diagonal error as well. For example, in our LRG analysis in section \ref{sec:lrgs} we choose $a_{\rm fac} = 1.1$ \cite{Ho:2012vy}.

Finally, we calculate the off-diagonal elements of the covariance matrix by preserving the structure of the OQE covariance matrix,
\bea
	\sigma^{2} \big( C_{\ell,\ell'}^{\alpha,\alpha} \big) & = & \frac{\sigma^{2} \big( C_{\ell,\ell',{\rm OQE}} \big)}{\sqrt{ \sigma^{2} \big( C_{\ell,{\rm OQE}} \big) \sigma^{2} \big( C_{\ell',{\rm OQE}} \big) }} \  \sqrt{\sigma^{2} \big( C_{\ell}^{\alpha,\alpha} \big) \sigma^{2} \big( C_{\ell'}^{\beta,\beta} \big)},
\label{eq:cov3} \\
	\sigma^{2} \big( C_{\ell}^{\alpha,\alpha} C_{\ell'}^{\beta,\beta} \big) & = & \frac{\sigma^{2} \big( C_{\ell,\ell',{\rm OQE}} \big)}{\sqrt{ \sigma^{2} \big( C_{\ell,{\rm OQE}} \big) \sigma^{2} \big( C_{\ell',{\rm OQE}} \big) }} \  \sqrt{\sigma^{2} \big( C_{\ell}^{\alpha,\alpha} C_{\ell}^{\beta,\beta} \big) \sigma^{2} \big( C_{\ell'}^{\alpha,\alpha} C_{\ell'}^{\beta,\beta} \big)}.
\label{eq:cov4}
\eea
This is a valid approximation since all redshift slices usually have similar OQE covariance structures (arising from the use of a common mask) \cite{Ho:2012vy,Ho:2013lda}. One can then choose any redshift slice to generate the ratio in the above equations.

We will now apply the method described here to simulated (section \ref{sec:sims}) and real (section \ref{sec:lrgs}) angular power spectra in the following sections.

\section{Simulation}
\label{sec:sims}

In this section we generate mock angular power spectra contaminated with known and unknown systematics in two redshift slices, and obtain estimates of the bias by discarding heavily contaminated bins. We choose redshift slices $0.50 \leq z \leq 0.55$ and $0.55 \leq z \leq 0.60$, labeled LRG2s and LRG3s (`s' standing for `simulation'), with redshift distributions identical to those for LRGs in SDSS-III DR8 \cite{Ho:2012vy,Ross:2011cz}. We generate auto- and cross-power spectra using the equations in section \ref{subsec:theorycl} with bias values of 2.0 and 2.2 and the non-linear fitting parameter $a$ set to zero in the two redshift slices, assuming a WMAP9 + SN $\Lambda$CDM cosmology. We then add mock systematic fields for stellar contamination, sky brightness, and seeing variations to both, the auto- and cross-power spectra. In order to generate the mock systematic fields, we use actual maps of these systematics and their cross-correlations with the observed galaxy density given in \cite{Ho:2012vy} and slightly exaggerate the effect of stars. Having maps of the systematics and their cross-correlations with galaxies one can calculate the various $\epsilon_{i}^{\alpha}(\ell)$ coefficients using eq.\ (\ref{eq:solveep}), with $u^{\alpha}(\ell) = 0$, and add in these systematics to obtain the observed angular power spectra from the theoretical power spectra. We subsequently correct for only sky brightness and seeing variations, leaving stars as an unknown systematic in the density fields.\footnote{We checked that on correcting for all three systematics we reproduce the theoretical power spectra as expected.}

\begin{figure}[!h]
\begin{center}
	\includegraphics[width=6.0in,angle=0]{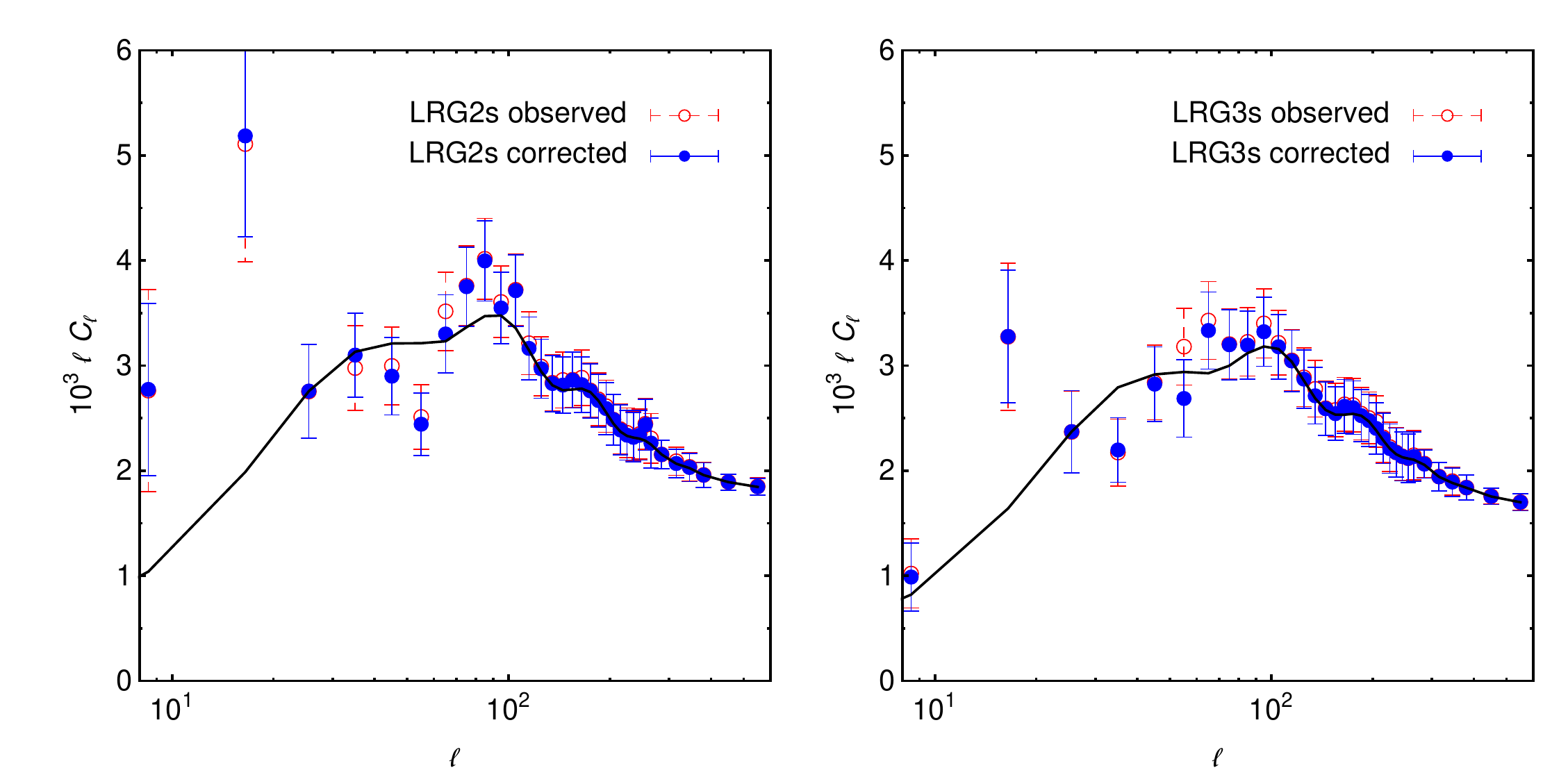}
	\includegraphics[width=3.0in,angle=0]{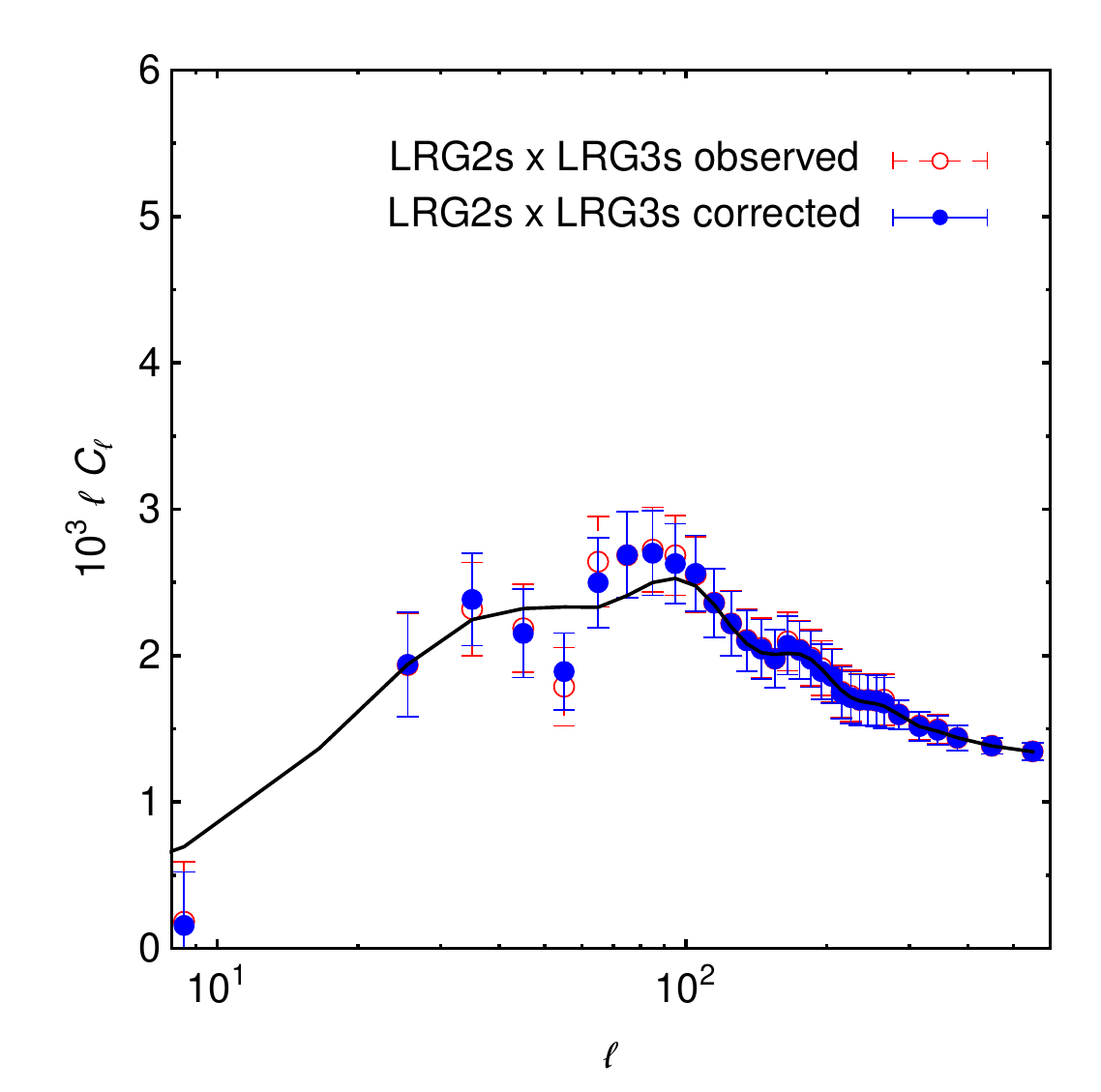}
	\caption{The auto- (top panels) and cross-power (bottom panel) spectra in simulated data for two redshift slices. The solid black curves are the original theoretical power spectra for bias values of 2.0 and 2.2 in redshift slices LRG2s and LRG3s, respectively. Red open circles with dashed ($1\sigma$) error bars are mock observed data points in which we have added three systematics, while blue filled circles with solid ($1\sigma$) error bars are data points corrected for only two systematics. The corrected cross-power is calculated analogous to the auto-power, using eq.\ (\ref{eq:crosspower}) with $U_{\ell}^{\alpha,\beta} = 0$.}
\label{fig1}
\end{center}
\end{figure}

In order to correct for known systematics (i.e. sky and seeing) we first solve eq.\ (\ref{eq:solveep}) for the weights $\epsilon_{i}^{\alpha}(\ell)$ in each $\ell$ bin, under the assumption that $u^{\alpha}(\ell) = 0$. We then use eq.\ (\ref{eq:autopower}) to obtain the corrected auto-power spectra in each redshift slice. In fig.\ \ref{fig1} we present the auto- and cross-power spectra in the two redshift slices before and after adding systematics. We also show the power spectra corrected for known systematics for comparison. Eqs.\ (\ref{eq:cov1}) - (\ref{eq:cov4}) further provide the full covariance matrix for the MCMC analysis, where for the smooth power spectra we use the observed spectra that include all three systematics and for the off-diagonal elements we use the structure of the OQE covariance matrix from one of the redshift slices in DR8.

To compute the errors in the cross-power (which we do not need in our analysis but we do show in fig.\ \ref{fig1}), we use a modified version of the following equation that assumes Gaussian errors (see, e.g., \cite{White:2008jy,Ross:2011zza}),
\bea
	\sigma^{2} \big( C_{\ell}^{\alpha,\beta} \big) & = & \frac{1}{f_{\rm sky}(2\ell + 1)} \bigg( \left( C_{\ell,{\rm smooth}}^{\alpha,\beta} \right)^{2} \nonumber \\
	& & \quad + \ \left( C_{\ell,{\rm smooth}}^{\alpha,\alpha} + N_{{\rm shot},\alpha} \right) \left( C_{\ell,{\rm smooth}}^{\beta,\beta} + N_{{\rm shot},\beta} \right) \bigg),
\label{eq:covcross1}
\eea
to include the correctional power due to systematics,
\bea
	\sigma^{2} \big( C_{\ell}^{\alpha,\beta} \big) & = & a^{2}_{\rm fac} \frac{1}{f_{\rm sky} \sum_{\ell = \ell_{\rm min}}^{\ell_{\rm max} - 1}(2\ell + 1)} \Bigg( \left( C_{\ell,{\rm smooth}}^{\alpha,\beta} \right)^{2} + \left( \Delta C_{\ell}^{\alpha,\beta} \right)^{2} \nonumber \\
	& & \quad \quad + \ \left( \sqrt{ \left( C_{\ell,{\rm smooth}}^{\alpha,\alpha} \right)^{2} + \Big( \Delta C_{\ell}^{\alpha,\alpha} \Big)^{2} } + N_{{\rm shot},\alpha} \right) \nonumber \\
	& & \quad \quad \quad \times \ \left( \sqrt{ \left( C_{\ell,{\rm smooth}}^{\beta,\beta} \right)^{2} + \Big( \Delta C_{\ell}^{\beta,\beta} \Big)^{2} } + N_{{\rm shot},\beta} \right) \Bigg).
\label{eq:covcross2}
\eea
This is the variance in the cross-power between redshift slices $\alpha$ and $\beta$, which includes an extra contribution from the auto-power, since the auto-power effectively acts like an extra source of noise when computing the error in the cross-power.

\begin{figure}[!h]
\begin{center}
	\includegraphics[width=3.in,angle=0]{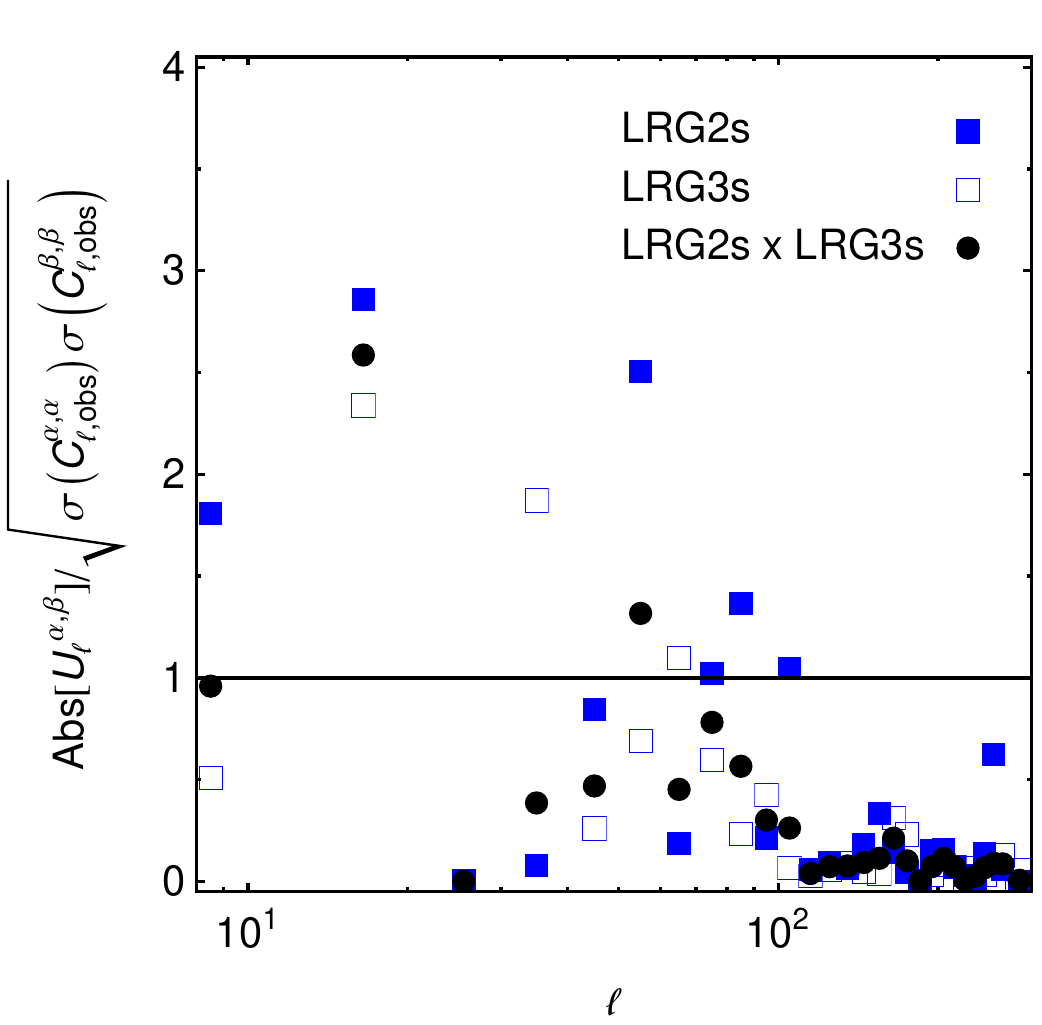}
	\caption{The absolute value of the unknown contamination $U^{\alpha,\beta}_{\ell}$ defined in eq.\ (\ref{eq:crosspower}) compared to the errors in the observed angular power spectra in the two redshift slices for simulated LRGs. Squares show the contamination in the auto-power spectra, i.e. ${\rm Abs}[U_{\ell}^{\alpha,\alpha}]/\sigma \big( C_{\ell, {\rm obs}}^{\alpha,\alpha} \big)$ while circles show the contamination in the cross-power. The solid line is the $1\sigma$-line.}
\label{fig2}
\end{center}
\end{figure}

Before we proceed with the MCMC analysis, in fig.\ \ref{fig2} we show the amount of unknown contamination (i.e. the contamination due to stars) in the auto- and cross-power spectra relative to the errors in the observed angular power spectrum in the two redshift slices. The quantity $U_{\ell}^{\alpha,\beta}$ defined in eq.\ (\ref{eq:crosspower}) is simply calculated as the difference between the corrected data points and the corresponding theoretical angular power spectra in fig.\ \ref{fig1}. Then fig.\ \ref{fig2} shows that in redshift slice LRG2s, bins at $\ell = 8.5, \ 16.5, \ 55, \ 75, \ 85, \ 105$ are contaminated at more than one sigma, while in redshift slice LRG3s, bins at $\ell = 16.5, \ 35, \ 65$ are similarly contaminated. The cross-power, however, is expected to pick up contaminations in bins at $\ell = 16.5, \ 55$. We will see below that this is in fact what we find in the MCMC analysis.

Now we perform the MCMC analysis described in section {\ref{subsec:mcmc}}. We choose a low-$\ell$ cutoff for the angular power spectrum in each redshift slice at $\ell_{\rm min} = 30$ since we expect lower multipoles to mostly be dominated by systematics. We also choose a high-$\ell$ cutoff, $\ell_{\rm max}$, corresponding to $k = 0.1 h \ {\rm Mpc}^{-1}$ (determined using $\Lambda{\rm CDM}$ cosmology, see table\ \ref{table1}) to avoid the strongly non-linear regime of the matter power spectrum. In addition to the bias and non-linear fitting parameter in each redshift slice, we also vary over the standard cosmological parameters $\big\{ \Omega_{b}h^{2}, \Omega_{\rm DM}h^{2}, \theta,\tau, n_{s}, \log A_{s}, A_{\rm SZ} \big\}$. Here $\Omega_{b}h^{2}$ is the physical baryon density, $\Omega_{\rm DM}h^{2}$ is the physical dark matter density, $\theta$ is the ratio of the sound horizon to the angular diameter distance at decoupling, $\tau$ is the reionization optical depth, $n_{s}$ is the scalar spectral index, $A_{s}$ is the amplitude of the primordial scalar curvature perturbations at $k =0.05 \ {\rm Mpc}^{-1}$, and $A_{\rm SZ}$ represents a Sunyaev-Zeldovich template normalization. We use flat priors for all parameters. This analysis yields the bias and $a$ parameters shown in the sixth and seventh columns of table\ \ref{table1}.

\begin{table}[!h]
\begin{center}
	\begin{tabular}{|c|c|c|c|c|c|c|c|c|}
		\hline
		Label & $z_{\rm mid}$ & Input & Input & $l_{\rm max}$ & $b_{1}$ & $10^{6}a$ & $b_{1}$ & $10^{6}a$ \\
		& & $b_{1}$ & $a$ & & & & ($1^{\rm st}$ it.) & ($1^{\rm st}$ it.) \\
		\hline
		LRG2s & 0.525 & 2.0 & 0 & 140 & $1.91^{+0.09}_{-0.09}$ & $2.97^{+2.71}_{-2.72}$ & $2.03^{+0.08}_{-0.09}$ & $-0.04^{+2.97}_{-2.92}$ \\
		LRG3s & 0.575 & 2.2 & 0 & 151 & $2.11^{+0.10}_{-0.09}$ & $2.02^{+2.33}_{-2.36}$ & $2.15^{+0.09}_{-0.10}$ & $1.31^{+2.44}_{-2.44}$ \\
		\hline
	\end{tabular}
\caption{The best-fit Gaussian bias and non-linear fitting parameter (with $1\sigma$ errors) in the two simulated redshift slices for LRGs, using WMAP9 + SN + simulated LRG data. In the sixth and seventh columns we have used all available $\ell$ bins in $30 \leq \ell \leq \ell_{\rm max}$, while in the next two columns we have used only those bins in $10 \leq \ell \leq \ell_{\rm max}$ that satisfy a $1\sigma$-cut on ${\cal U}^{\alpha,\beta}_{\ell}$.}
\label{table1}
\end{center}
\end{table}

\begin{figure}[!h]
\begin{center}
	\includegraphics[width=3.in,angle=0]{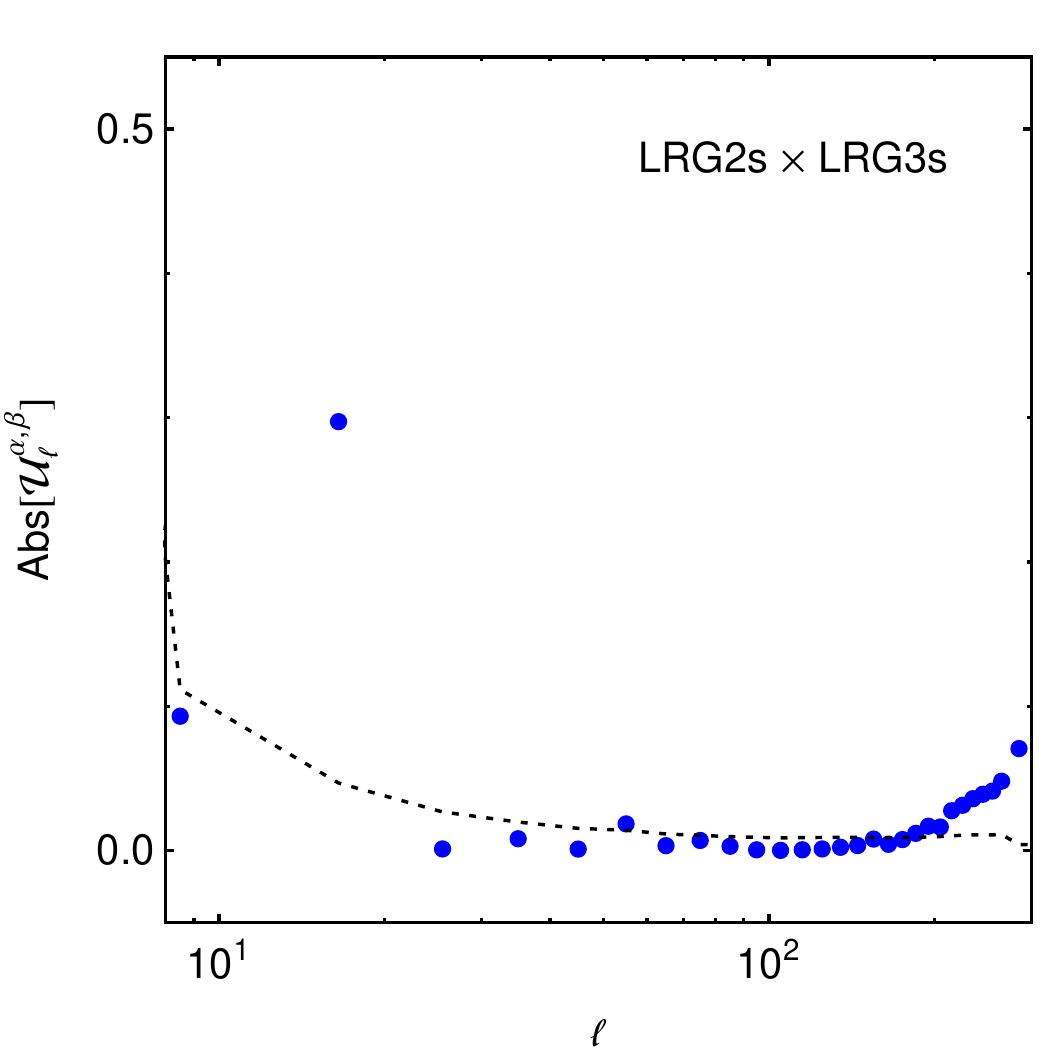}
	\caption{The absolute value of the unknown contamination coefficient ${\cal U}^{\alpha,\beta}_{\ell}$ defined in eq.\ (\ref{eq:defineul}) for simulated LRGs (filled circles). The dotted line shows the absolute value of the $1\sigma$-cut --- we drop all bins that lie above this cut. The upturn in the value of ${\cal U}^{\alpha,\beta}_{\ell}$ at high $\ell$ comes from the cross-power being dominated by the non-linear fitting parameters on these scales.}
\label{fig3}
\end{center}
\end{figure}

Next we use the best-fit bias and $a$ parameters to obtain the theoretical cross-power spectrum between the two redshift slices. As noted earlier, we calculate the cross-redshift distribution needed here as an overlap of the distribution in the two slices. We also add $\sqrt{|a^{\alpha}| \ |a^{\beta}|}$ to the theoretical cross-power, to take into account any non-linear contributions in the region with which we are concerned. Comparing this `true' power spectrum with the observed cross-power and using the weights $\epsilon_{i}^{\alpha}(\ell)$ obtained earlier, we calculate $U^{\alpha,\beta}_{\ell}$ using eq.\ (\ref{eq:crosspower}) in each $\ell$ bin. Finally we determine the unknown contamination coefficient ${\cal U}^{\alpha,\beta}_{\ell}$ using eq.\ (\ref{eq:defineul}). We show the absolute value of this quantity with the $1\sigma$-cut that we apply in fig.\ \ref{fig3} and note that bins with a contamination greater than one sigma do indeed correspond to those shown earlier in the cross-power in fig.\ \ref{fig2}.

We now drop all $\ell$ bins in the auto-power of each of the two redshift slices, for which the unknown contamination coefficient lies outside the $1\sigma$-cut. Using only the remaining bins in $10 \leq \ell \leq \ell_{\rm max}$ and the corresponding rows and columns of the full covariance matrix obtained earlier, we perform an MCMC analysis on the standard cosmological parameters, the bias, and the $a$ parameter in each redshift slice. This exercise yields the bias and $a$ values denoted as `$1^{\rm st}$ it.' (First iteration) in table\ \ref{table1}. The bias in the redshift slice LRG2s lies slightly outside the one sigma bound of the previous estimate. When we used these new estimates of the bias to determine which bins to drop, however, we ended up with the same bins in $10 \leq \ell \leq \ell_{\rm max}$ as the first iteration. These values of the bias are therefore the final values obtained on applying our method. As one can see, estimates of the bias are now closer to the original values of 2.0 and 2.2. We also find that this fit leads to a reduction in $\chi^2$ and thus represents an improved fit to the data.

\begin{figure}[!h]
\begin{center}
	\includegraphics[width=6.0in,angle=0]{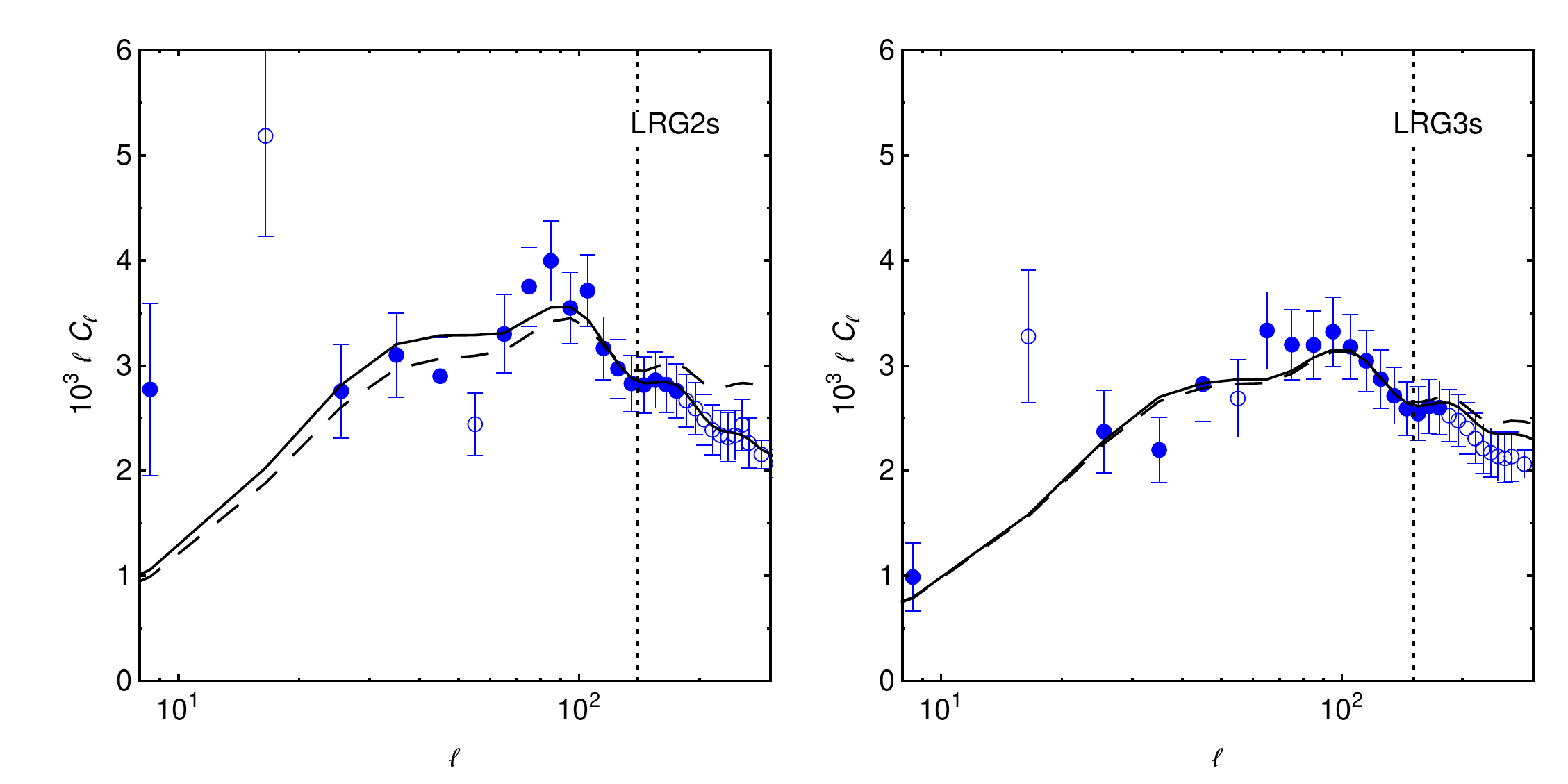}
	\caption{The angular power spectrum in the two simulated redshift slices of LRGs. Open circles with ($1\sigma$) error bars represent data points that are dropped due to large unknown systematics, as determined using the cross-power spectra. Filled circles, on the other hand, are data points that are not dominated with unknown systematics. The vertical dotted line shows $\ell_{\rm max}$. The dashed curve is the theoretical power spectrum obtained using the best-fit bias parameters that correspond to the MCMC analysis which uses all data points in $30 \leq \ell \leq \ell_{\rm max}$. The solid curve represents the MCMC analysis which uses only filled circles in $10 \leq \ell \leq \ell_{\rm max}$. In both theoretical curves we have added in the corresponding best-fit $a$ parameter to all $C_{\ell}$s; this causes the upturn at large $\ell$.}
\label{fig4}
\end{center}
\end{figure}

In fig.\ \ref{fig4} we present the angular power spectrum corrected for known systematics and mark the bins that we drop in each redshift slice. We also show theoretical curves obtained using best-fit values of the bias and the $a$ parameter before and after dropping bins. According to the method presented here we exclude bins from each of the two redshift slices that have a significantly contaminated cross-correlation, as we do not know which of the two redshift slices is responsible for this contamination. For this reason, there are some $\ell$ bins in the auto-power spectra of each redshift slice that may not appear to be contaminated but are still dropped.

It is important to note that the theoretically calculated `true' cross-power spectrum depends on the assumed background cosmological model. There is a possibility therefore that the bins that are discarded could be showing a real cosmological signal of interest. In order to assess this effect, we repeat the analysis of this section with a different background cosmology. In particular, we consider the case of non-zero primordial non-Gaussianity. As was first reported in \cite{Dalal:2007cu}, the halo bias includes a scale-dependent term given by
\bea
	\Delta b(M,k,z,f_{\rm NL}) & = & 3f_{\rm NL}[b_{1}(M,z) - p]\delta_{c} \frac{\Omega_{m}H_{0}^{2}}{k^{2}T(k)D(z)}
\label{eq:deltabstd}
\eea
in the presence of primordial (local) non-Gaussianity, usually parameterized by the parameter $f_{\rm NL}$. Here, $\delta_{c} \approx 1.686$ denotes the critical density for spherical collapse, $\Omega_{m}$ is the present day matter density, $H_{0}$ is the Hubble constant, $T(k)$ is the matter transfer function normalized to unity as $k \rightarrow 0$, and $D(z)$ is the linear growth function normalized to $(1+z)^{-1}$ in the matter-dominated era. The parameter $p$ ranges from unity for objects that populate all halos equally to $1.6$ for objects that populate only recently merged halos \cite{Slosar:2008hx}. For LRGs we can set $p$ to unity. We generate auto- and cross-power spectra as before, assuming a WMAP9 + SN $\Lambda$CDM cosmology, but in addition setting $f_{\rm NL} = 25$. We then add in three mock systematic fields and correct for two of them, leaving the third as an unknown systematic. We subsequently perform an MCMC analysis assuming that $f_{\rm NL} = 0$, expecting to find more low-$\ell$ bins to be contaminated with unknown systematics, since the effect of non-zero $f_{\rm NL}$ is more predominant on large scales. Since the effect of systematics is large compared to the actual cosmological signal of a small non-zero $f_{\rm NL}$, however, we find results very similar to what we found earlier, with the method dropping the same bins as before. We do find, however, slightly higher estimates of the Gaussian bias since $f_{\rm NL} > 0$ increases the power on large scales. With more precise data in the future it may be useful to implement the method presented here for each parameter space being explored, for instance including $f_{\rm NL}$ in the MCMC analysis that determines which bins to discard and using the resulting best-fit value of $f_{\rm NL}$ to calculate the true cross-power.

\section{A case study: SDSS-III LRGs}
\label{sec:lrgs}

Let us now apply the method discussed in sections \ref{sec:systematics} and \ref{sec:sims} to actual data from SDSS-I, II and III \cite{Gunn:1998vh,Gunn:2006tw,Bolton:2012hz,Smee:2012wd}. We use LRGs from the SDSS-III DR8 sample described in \cite{Ho:2012vy,Ross:2011cz}. We refer the reader to these papers for details and highlight some of the main properties of the sample here. The data set spans $\sim 11,000$ square degrees of the sky and probes a volume of $\sim 3h^{-3} \ {\rm Gpc^{3}}$. We focus on the approximately stellar mass-limited CMASS sample of luminous galaxies, that follows the CMASS galaxy selection detailed in \cite{White:2010ed}. Photometric redshifts and the probability that an object is a galaxy are obtained using a training sample of 112,778 BOSS CMASS spectra. The final photometric redshift catalog consists of a total of 872,921 luminous galaxies in the redshift range $0.45 \leq z \leq 0.65$, divided into four photometric bins (labeled LRG1 through LRG4) --- $z = 0.45-0.50$, $0.50-0.55$, $0.55-0.60$, and $0.60-0.65$, with the effective number of galaxies in each bin being $214971$, $258736$, $248895$, and $150319$, respectively. The calculation of the angular power spectra in the four redshift bins uses the OQE method outlined in section \ref{subsec:oqe}.

We first take the output of the OQE and correct for dominant known systematics, which include stellar contamination, sky brightness, and seeing variations. The maps of these systematics and their cross-correlations with the observed galaxy density field were given in \cite{Ho:2012vy}, so we will not repeat them here. As before, we correct for these systematics assuming that there are no unknown systematics, and also obtain the full covariance matrix that is used in the MCMC analysis.

\begin{figure}[!h]
\begin{center}
	\includegraphics[width=5.0in,angle=0]{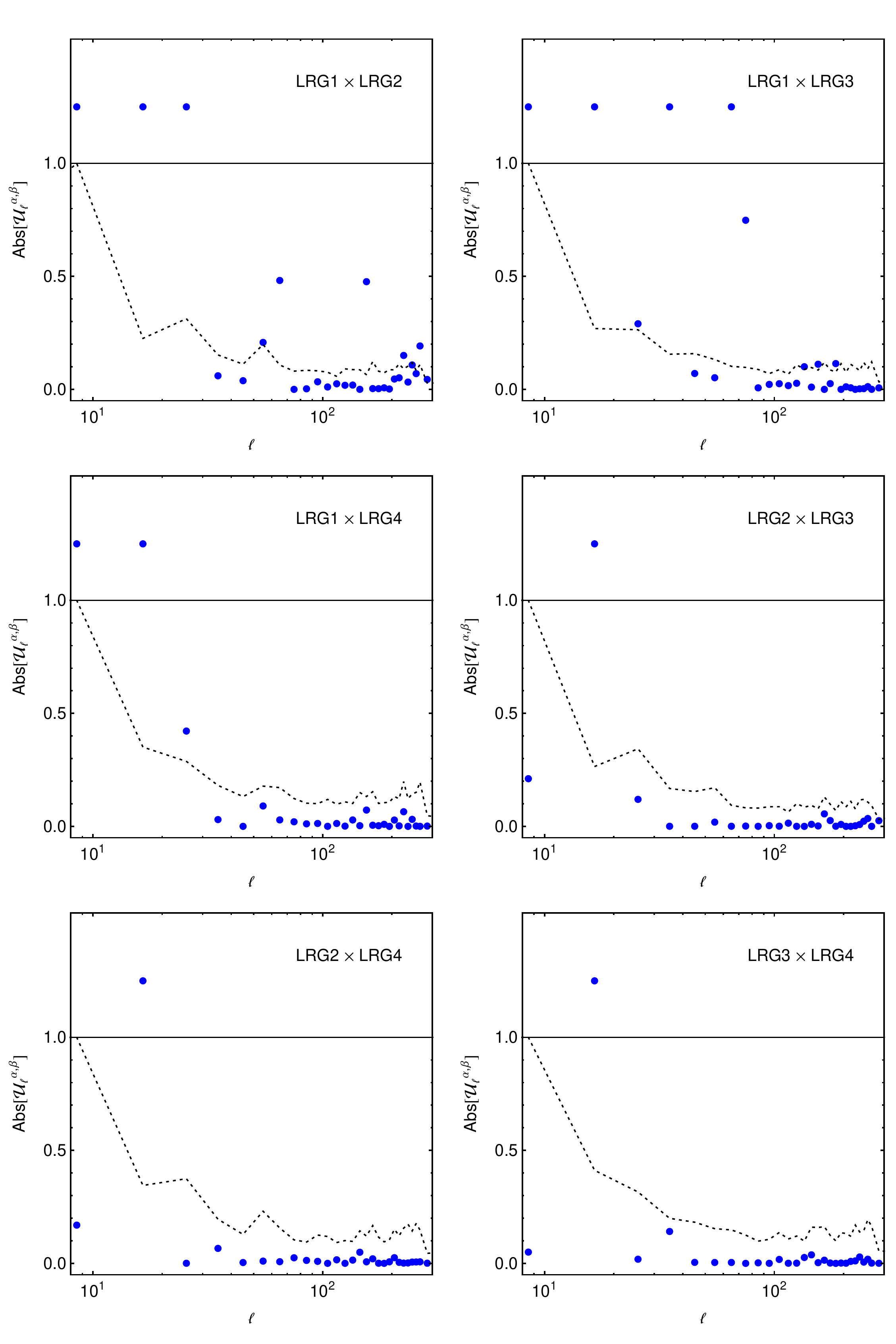}
	\caption{The absolute value of the unknown contamination coefficient ${\cal U}^{\alpha,\beta}_{\ell}$ defined in eq.\ (\ref{eq:defineul}) for SDSS-III DR8 LRGs (filled circles). The dotted line shows the absolute value of the $3\sigma$-cut --- we drop all bins in the corresponding auto-power spectra that lie above this cut. The points that lie above the solid line have ${\rm Abs}\big[{\cal U}^{\alpha,\beta}_{\ell}\big] > 1$.}
\label{fig5}
\end{center}
\end{figure}

We perform the MCMC analysis described in section {\ref{subsec:mcmc}}, choosing a low-$\ell$ cutoff for the angular power spectrum in each redshift slice at $\ell_{\rm min} = 30$ and a high-$\ell$ cutoff, $\ell_{\rm max}$, corresponding to $k = 0.1 h \ {\rm Mpc}^{-1}$ (see table\ \ref{table2}). In addition to the bias and non-linear fitting parameter in each redshift slice, we also vary over the standard cosmological parameters $\big\{ \Omega_{b}h^{2}, \Omega_{\rm DM}h^{2}, \theta,\tau, n_{s}, \log A_{s}, A_{\rm SZ} \big\}$. The resulting bias and $a$ parameters obtained are given in the fourth and fifth columns of table\ \ref{table2}. Using the best-fit bias and $a$ parameters we calculate the theoretical cross-power spectra between the four redshift slices, adding $\sqrt{|a^{\alpha}| \ |a^{\beta}|}$ to the cross-power between consecutive redshift slices. Comparing these `true' power spectra with the observed cross-power and using the weights $\epsilon_{i}^{\alpha}(\ell)$ obtained earlier (under the assumption of no unknown systematics), we calculate the unknown contamination coefficient ${\cal U}^{\alpha,\beta}_{\ell}$. We show the absolute value of this quantity along with the $3\sigma$-cut that we apply in fig.\ \ref{fig5}. We choose a more conservative $3\sigma$-cut as we are now dealing with real data.

\begin{table}[!h]
\begin{center}
	\begin{tabular}{|c|c|c|c|c|c|c|}
		\hline
		Label & $z_{\rm mid}$ & $l_{\rm max}$ & $b_{1}$ & $10^{6}a$ & $b_{1}$ & $10^{6}a$ \\
		& & & & & ($1^{\rm st}$ it.) & ($1^{\rm st}$ it.) \\
		\hline
		LRG1 & 0.475 & 128 & $1.89^{+0.09}_{-0.10}$ & $4.93^{+4.38}_{-4.27}$ & $1.96^{+0.14}_{-0.14}$ & $2.86^{+5.78}_{-5.79}$ \\
		LRG2 & 0.525 & 140 & $1.88^{+0.09}_{-0.10}$ & $6.32^{+2.90}_{-2.88}$ & $1.97^{+0.11}_{-0.10}$ & $3.83^{+3.32}_{-3.30}$ \\
		LRG3 & 0.575 & 151 & $2.16^{+0.10}_{-0.09}$ & $0.74^{+2.48}_{-2.45}$ & $2.09^{+0.12}_{-0.13}$ & $2.14^{+3.10}_{-3.08}$ \\
		LRG4 & 0.625 & 162 & $2.26^{+0.11}_{-0.11}$ & $1.40^{+2.31}_{-2.32}$ & $2.26^{+0.11}_{-0.12}$ & $1.44^{+2.35}_{-2.35}$ \\
		\hline
	\end{tabular}
\caption{The best-fit Gaussian bias and non-linear fitting parameter (with $1\sigma$ errors) in the four redshift slices for LRGs, using WMAP9 + SN + DR8 (LRG) data. In the fourth and fifth columns we have used all available $\ell$ bins in $30 \leq \ell \leq \ell_{\rm max}$, while in the next two columns we have used only those bins in $10 \leq \ell \leq \ell_{\rm max}$ that satisfy a $3\sigma$-cut on ${\cal U}^{\alpha,\beta}_{\ell}$.}
\label{table2}
\end{center}
\end{table}

We now drop all $\ell$ bins in the auto-power of each redshift slice, whose unknown contamination in the cross-power with some other redshift slice lies outside the corresponding $3\sigma$-cut. Again, using only the remaining bins in $10 \leq \ell \leq \ell_{\rm max}$ and the corresponding rows and columns of the full covariance matrix obtained earlier, we perform an MCMC analysis on the standard cosmological parameters, the bias, and the $a$ parameter in each redshift slice. This yields the bias and $a$ values given in the last two columns of table\ \ref{table2}. Values of the bias lie within one sigma of the previous bias estimates, so these are the final values to use.

\begin{figure}[!h]
\begin{center}
	\includegraphics[width=6.0in,angle=0]{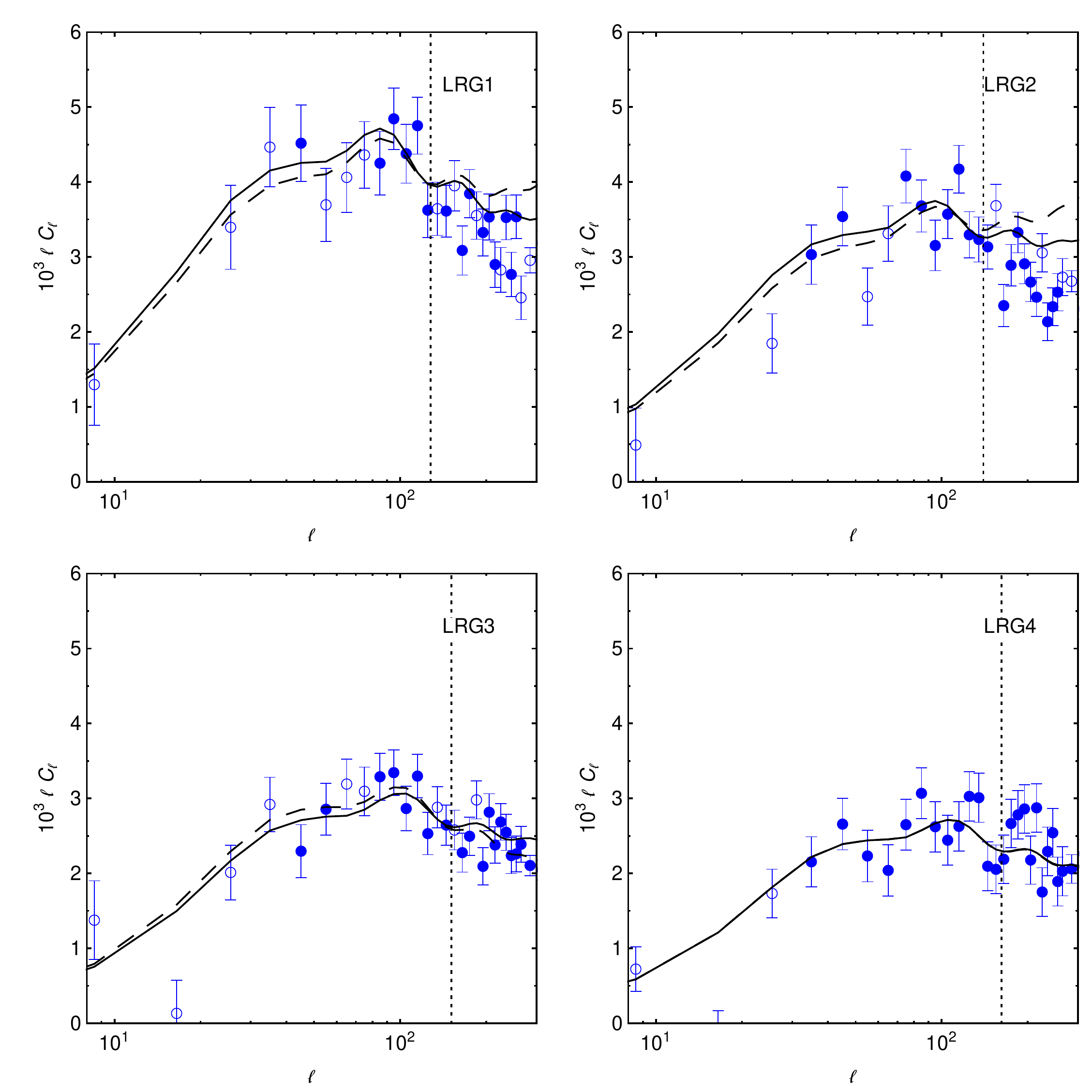}
	\caption{The angular power spectrum in the four redshift slices of LRGs. Open circles with ($1\sigma$) error bars represent data points that are dropped due to large unknown systematics, as determined using the cross-power spectra. Filled circles, on the other hand, are data points that are not dominated with unknown systematics. The vertical dotted line shows $\ell_{\rm max}$. The dashed curve is the theoretical power spectrum obtained using the best-fit bias parameters that correspond to the MCMC analysis which uses all data points in $30 \leq \ell \leq \ell_{\rm max}$. The solid curve represents the MCMC analysis which uses only filled circles in $10 \leq \ell \leq \ell_{\rm max}$. In both theoretical curves we have added in the corresponding best-fit non-linear fitting parameter $a$ to all $C_{\ell}$s; this causes the upturn at large $\ell$.}
\label{fig6}
\end{center}
\end{figure}

In fig.\ \ref{fig6} we present the measured angular power spectrum in each redshift slice after correcting for known systematics and mark the bins that are dropped based on the method discussed above. We also display theoretical curves\footnote{The theoretical curves shown in fig.\ \ref{fig6} differ from those used in calculating the likelihood by the effect of the survey window function.} obtained using best-fit values for the bias and $a$ before and after dropping bins. In each redshift slice there are some bins that do not appear to be contaminated but are still dropped as their cross-power with another redshift slice is significantly contaminated.

To see how this method changes the estimates of other cosmological parameters, we examine a specific example of primordial non-Gaussianity. We add the term given in eq.\ (\ref{eq:deltabstd}) to the bias $b_{1}$ and introduce an extra parameter $f_{\rm NL}$ to the MCMC analysis. We do {\it not} include $f_{\rm NL}$ in the MCMC method to determine which bins to drop, but rather only include it as a parameter to fit {\it after} determining which bins to drop assuming $f_{\rm NL} = 0$. In order to constrain $f_{\rm NL}$ we set the (Gaussian) bias $b_{1}$ and the non-linear fitting parameter $a$ to their corresponding best-fit values in each redshift slice, and perform an MCMC analysis over the standard cosmological parameters and $f_{\rm NL}$. Using all $\ell$ bins in $30 \leq \ell \leq \ell_{\rm max}$ we find that $f_{\rm NL} = -5^{+62}_{-62} \ (1\sigma {\rm \ error})$, using instead all bins in $10 \leq \ell \leq \ell_{\rm max}$ we obtain a tighter constraint of $f_{\rm NL} = -77^{+47}_{-46} \ (1\sigma {\rm \ error})$, and finally using only those bins in $10 \leq \ell \leq \ell_{\rm max}$ that satisfy the $3\sigma$-cut we obtain $f_{\rm NL} = -17^{+68}_{-68} \ (1\sigma {\rm \ error})$. For data at higher redshifts (such as quasars) we expect the variation in parameter estimation to be even more significant, and therefore propose that one should use a method such as that discussed in this paper to exclude significantly contaminated bins in photometric LSS surveys (for instance, see \cite{Ho:2013lda}).

\section{Discussion}
\label{sec:discussion}

Photometric surveys are currently limited by systematics. In order to correctly use data from current and upcoming surveys, it is important to understand and account for various sources of contamination that affect photometric samples. This issue has motivated works such as \cite{Pullen:2012rd,Giannantonio:2013uqa,Leistedt:2013gfa}. In \cite{Pullen:2012rd}, the authors demonstrated that auto-correlations of quasars in the SDSS DR6 sample \cite{AdelmanMcCarthy:2007aa,Richards:2008eq} are unfit to constrain primordial non-Gaussianity. They constructed templates for various potential systematic effects and mode-projected them from the angular cross-power spectra. Although this improved the significance of the cross-correlation measurement, they concluded that other systematics are still contaminating the SDSS photometric quasar sample. In \cite{Leistedt:2013gfa}, the authors used improved sky masks and mode projection to further reduce contamination levels, and in \cite{Giannantonio:2013uqa}, the authors obtained improved cosmological parameter estimates by using cross-correlations between different surveys.

In this paper we have adopted a different approach and have developed a method to exclude bins in the angular power spectrum that are significantly contaminated by unknown systematics. For this purpose, we use cross-correlations between different redshift slices and define an unknown contamination parameter to estimate the contribution from unknown systematics in the auto-power spectra. This allows one to selectively drop bins that lie outside a specified contamination tolerance. We apply our method on simulated data for LRGs in the SDSS to demonstrate that it does improve estimates of the bias in each redshift slice. We further apply the method to real LRGs in the SDSS-III DR8 sample as a case study. This also allows a comparison, as an example of cosmological parameter estimation, of constraints on primordial non-Gaussianity before and after applying our method.

In the analysis presented in this paper, we assume a certain model for non-linear structure formation --- in particular, we use the HaloFit prescription and model non-linearities with an extra $a$ parameter. Although we cut the analysis at $k = 0.1 h \ {\rm Mpc}^{-1}$, we expect some amount of unknown contamination to be attributed to a lack of theoretical understanding of the matter power spectrum in the non-linear regime. We would also like to emphasize that an imperfect modeling of known systematic fields, for instance a more complex relationship compared to the simple linear one we use between systematics and the observed density, is also expected to contribute to unknown systematics. Further, as discussed in section\ \ref{sec:sims} the true cross-power depends on the assumed background cosmological model; one could thus implement the method presented here for each cosmological parameter space being studied.

It is also worthwhile to note that the method discussed here provides an upper estimate of the errors since it drops bins from {\it both} redshift slices whose cross-correlation is significantly contaminated. It may be possible to use different combinations of the auto- and cross-power spectra and additionally cross-correlations between photometric and spectroscopic measurements to derive less aggressive algorithms. Further, it would be useful to generalize such a method for spectroscopic surveys. These issues will be dealt with in future work.


\acknowledgments It is a pleasure to thank Dragan Huterer and An\v{z}e Slosar for many useful comments on this paper. We also thank Rachel Mandelbaum, Hiranya Peiris, Anthony Pullen, and Uro\v{s} Seljak for helpful discussions. N. A. is supported by the McWilliams fellowship of the Bruce and Astrid McWilliams Center for Cosmology. N. A. also acknowledges support by the New Frontiers in Astronomy and Cosmology program at the John Templeton Foundation.

Funding for SDSS-III has been provided by the Alfred P. Sloan Foundation, the Participating Institutions, the National Science Foundation, and the U.S. Department of Energy Office of Science. The SDSS-III web site is http://www.sdss3.org/.

SDSS-III is managed by the Astrophysical Research Consortium for the Participating Institutions of the SDSS-III Collaboration including the University of Arizona, the Brazilian Participation Group, Brookhaven National Laboratory, University of Cambridge, Carnegie Mellon University, University of Florida, the French Participation Group, the German Participation Group, Harvard University, the Instituto de Astrofisica de Canarias, the Michigan State/Notre Dame/JINA Participation Group, Johns Hopkins University, Lawrence Berkeley National Laboratory, Max Planck Institute for Astrophysics, Max Planck Institute for Extraterrestrial Physics, New Mexico State University, New York University, Ohio State University, Pennsylvania State University, University of Portsmouth, Princeton University, the Spanish Participation Group, University of Tokyo, University of Utah, Vanderbilt University, University of Virginia, University of Washington, and Yale University.


\bibliography{references}
\bibliographystyle{JHEP}

\end{document}